\newcommand{\D}{\mathcal{D}}

\newcommand{\N}{\mathcal{N}}

\def\balpha{{\boldsymbol{\alpha}}}
\def\bbeta{{\boldsymbol{\beta}}}

\documentclass[epj]{svjour}
\usepackage{graphics}
\usepackage{booktabs}
\usepackage{amsmath, amssymb}
\usepackage{url}
\usepackage{xcolor}
\usepackage{subcaption}
\usepackage[ruled,vlined]{algorithm2e}
\usepackage[normalem]{ulem}

\definecolor{amethyst}{rgb}{0.6, 0.4, 0.8}

\definecolor{vermilion}{rgb}{0.89, 0.26, 0.2}

\begin{document}
  
\title{On Focusing Statistical Power for Searches and Measurements in Particle Physics}
\author{James Carzon\inst{1} \and Aishik Ghosh\inst{2,3,4} \and Rafael Izbicki\inst{4} \and Ann B. Lee\inst{1} \and Luca Masserano\inst{1} \and Daniel Whiteson\inst{2}
}                     
%
%
\institute{Department of Statistics and Data Science, Carnegie Mellon University, Pittsburgh, Pennsylvania, USA, \email{jcarzon@andrew.cmu.edu} \and Department of Physics and Astronomy, University\ of\
  California, Irvine, California, USA \and Physics Division, Lawrence Berkeley National Laboratory, Berkeley, California, USA \and School of Physics, Georgia Institute of Technology, Georgia, USA, \email{aishikghosh@gatech.edu} \and Department of Statistics, Federal University of São Carlos, Brazil}
\date{Received: date / Revised version: date}
%
\abstract{
Particle physics experiments rely on the (generalised) likelihood ratio test (LRT) for searches and measurements, which consist of composite hypothesis tests. However, this test is not guaranteed to be optimal, as the Neyman-Pearson lemma pertains only to simple hypothesis tests. Any choice of test statistic thus implicitly determines how statistical power varies across the parameter space. An improvement in the core statistical testing methodology for general settings with composite tests would have widespread ramifications across experiments. We discuss an alternate test statistic that provides the data analyzer an ability to focus the power of the test on physics-motivated regions of the parameter space. We demonstrate the improvement from this technique compared to the LRT on a Higgs $\rightarrow\tau\tau$ dataset simulated by the ATLAS experiment and a dark matter dataset inspired by the LZ experiment. We also employ machine learning to efficiently calibrate critical values for a family of tests, which are then inverted to obtain statistically valid confidence intervals.
%
} 
\maketitle
\section{Introduction}
\label{intro}
A major theme in high-energy physics is the search for new phenomena and the desire to precisely measure known physics parameters. The design of such experiments can be expensive, and it can take decades to achieve the required sensitivity to weak or rare signals in an ocean of uninteresting backgrounds. A {\em search} in this context concludes when the existence of a particle has been confidently established after analyzing enough collisions, usually by rejecting the default hypothesis that the observed excess in a given sample would occur by chance, as with the Higgs boson discovery~\cite{ATLAS:2012yve,CMS:2012qbp}. Alternatively, the search may conclude by confidently ruling out the existence of a certain class of particles, as with several supersymmetry and dark matter searches~\cite{LZ:2024zvo,Kwon:2025fgb}. Similarly, a {\em measurement} is reported in the form of a confidence interval on a key property of the particle --- a smaller interval means a more precise measurement. 

Experimental particle physicists have almost universally relied on the generalized likelihood ratio test (LRT) for such studies \cite{cousins_lectures_2024,cranmer_practical_2015}. The LRT is optimal for simple-vs-simple hypothesis tests (Neyman-Pearson's lemma, \cite{neyman_ix_1933}), and also has convenient asymptotic properties \cite{wilks_large-sample_1938,Cowan:2010js}.

However, searches and measurements are typically performed as {\em composite} hypothesis tests. For instance, the measurement of a signal strength parameter for a potential new phenomenon or the mass of a known particle generally involves testing a default conjecture, such as $H_0: \mu = \mu_0$, against a composite alternative hypothesis, such as $H_1:\mu \neq \mu_0$, which encompasses a range of possible parameter values rather than a single one.  In these cases, the Neyman–Pearson lemma no longer applies: the most powerful test may vary depending on the specific value $\mu_1$ within the alternative. In fact, for general settings, there is no single test that is  {\em simultaneously} most powerful across all parameter values in the composite alternative. Effectively, this means that any choice of test statistic is implicitly a choice for how the statistical power of the test varies over the parameter space; no test will be the most powerful everywhere. 

In this Article, we present an alternative to LRT that allows the data analyst to focus the power of a test on specific regions of the parameter space --- without assuming simple hypotheses or large sample sizes, while maintaining validity of confidence intervals. Our proposed {\em focused test statistic} (FTS) compares the null hypothesis to a weighted average of all alternative hypotheses, where the weighting is defined by a physics-motivated ``focus function'' (which is not determined by the data) that increases the test's sensitivity to the regions of the parameter space of most interest to physicists. We show that the FTS outperforms the LRT statistic for a collider physics example and a dark matter search example. The FTS has Bayesian roots: in Bayesian testing, it is known as the Bayes factor \cite{kass_bayes_1995}. However, in this work we use the FTS in a purely frequentist framework with theoretical guarantees of validity (i.e. coverage) and power (see Section~\ref{sec:statistical_properties}). The focus function may be determined through optimization studies and crucially---unlike Bayesian credible regions---our confidence intervals remain valid even when the focus is placed away from the true parameter value. The practice of using the Bayes factor as a frequentist test statistic to increase power was originally proposed by I.J. Good who referred to the practice as ``the Bayes/non-Bayes comprise''~\cite{good_bayesnon-bayes_1992}.

For our two case studies, we achieve significant gains using FTS in regimes with small sample sizes and low signal fractions, which is precisely where the most interesting physics can be found. Our method constructs confidence intervals by inverting hypothesis tests, following the classical construction \cite{neyman_outline_1937,casella_statistical_2024}. This construction is common in particle physics, where confidence intervals for complex models and datasets are often obtained through computationally expensive Monte Carlo pseudo-experiments~\cite{ATLAS:2012gfw,cranmer_practical_2015}. A prominent example is the Feldman--Cousins unified construction~\cite{Feldman:1997qc}, which builds a Neyman confidence belt using a likelihood-ratio ordering of the sample space and gives a unified treatment of upper limits and two-sided confidence intervals. Our contribution is to retain this inversion-based perspective while replacing the traditional Monte Carlo belt construction with a faster machine learning-based procedure \cite{pmlr-v119-dalmasso20a,dalmasso_likelihood-free_2024,masserano_simulator-based_2023}. This technique applies to either test statistic and does not require large sample sizes, thereby enabling the construction of valid confidence intervals in a range of dark matter, neutrino or collider experiments where local asymptotics cannot be assumed~\cite{cranmer_practical_2015,NOvA:2017abs,ATLAS:2024jry,LZ:2025xkj}.

As a drop-in replacement for the LRT statistic, FTS can be applied to any existing particle physics analysis and holds the potential to improve sensitivity in key measurements across experiments, such as neutrino oscillation parameters, Higgs self-coupling, Higgs width, effective field theory parameters and also searches, such as for dark matter.

This Article is organized as follows: Section~\ref{sec:StatisticalSetup} introduces the statistical setup. Section~\ref{sec:ExperimentsResults} compares the two test statistics for a Higgs measurement and a dark matter search problem. Section~\ref{sec:Conclusion} discusses our main results.

\section{Statistical Setup}
\label{sec:StatisticalSetup}

\subsection{How do we construct confidence intervals from tests?}\label{sec:construction}

Searches and measurements in particle physics are typically performed as hypothesis tests on the parameter $\mu$ of the likelihood $p(\mathcal{D}; \mu)$ of datum $\mathcal{D}$ implied by the physics processes under study for some parameter space $\Theta$. (See Appendix~\ref{sec:experimental_setup} for a detailed description of the experimental setup). For example, the {\em generalized LRT} rejects the null hypothesis $H_0: \mu=\mu_0$ (default conjecture on, e.g., the strength $\mu$ of a signal physics process) in favor of the alternative hypothesis $H_1: \mu \neq \mu_0$  when the generalized likelihood ratio statistic (LRS)
\begin{equation} \label{eq:LRS}
T(\D; \mu_0) = -2\log\left( \frac{p(\D; \mu_0)}{\sup_{\mu \in \Theta} p(\D; \mu)} \right)
\end{equation}
is large or above some ``critical value'' $C_{\mu_0}$, which is chosen so that the proportion of wrongly rejected events (``false positives'') do not exceed some pre-set significance level $\alpha$. These hypothesis tests can then also be converted into an interval estimate of $\mu$ by inverting the hypothesis tests \cite{neyman_outline_1937}, where a $(1-\alpha)100$\% {\em confidence interval} $I(\mathcal{D})$ of $\mu$ is defined by the set of all parameter values that passed the level-$\alpha$ tests:
\begin{equation} \label{eq:CI}
    I(\mathcal{D})=\{\mu_0 \in \Theta \mid T(\mathcal{D};\mu_0) < C_{\mu_0} \}.
\end{equation}
See Figure~\ref{fig:test_statistics} (top) for a depiction of the construction, and Appendix~\ref{sec:neyman_and_confidence_belts} for a discussion on parallels with the classical ``confidence belt'' figure \cite{Feldman:1997qc,neyman_outline_1937,clopper_use_nodate}. The statistical efficiency or ``power'' of the procedure (which determines how tight the intervals are) depends on the intricacies of the likelihood and the choice of the test statistic $T$. The computational efficiency of the procedure depends on how fast we can compute the test statistic $T(\mathcal{D};\mu_0)$ and the critical values $C_{\mu_0}$ as a function of $\mu_0$. In practice, $\Theta$ is a discrete but high resolution grid of values for $\mu_0$. Furthermore, to avoid boundary effects, we replace $\Theta$ in the denominator of Equation~\ref{eq:LRS} by an extended parameter space $\widetilde{\Theta}$, which is defined as the parameter space $\Theta$ padded with one unit at both ends; see Appendix~\ref{sec:implementation} for details.

\subsection{How do we cheaply compute critical values?}\label{sec:quantile_regression}

Data analyses in collider physics often rely on asymptotic properties (such as Wilks' theorem \cite{wilks_large-sample_1938,Cowan:2010js}) to cheaply estimate LRS confidence intervals, but asymptotics do not hold in small sample regimes. In this Article, we propose a machine learning (ML) approach to quickly estimate how $C_{\mu_0}$ might vary with $\mu_0$; see Algorithm~\ref{alg:estimate_cutoffs} in Appendix~\ref{sec:critical_values} for details of a procedure that performs a quantile regression (QR) of $T$ on $\mu_0$. An ML-based approach allows the data analyst to (i) take advantage of alternative test statistics (such as the proposed FTS defined below in Eq.~\ref{eq:FTS}),  (ii) explore challenging small-sample regimes, and  (iii)  extend HEP analyses to higher-dimensional parameter spaces (such as in effective field theory) or to unbinned analyses, where a Monte Carlo (MC) approach (e.g., \cite[Sec.~5.4]{cranmer_practical_2015}) is not computationally feasible.

\subsection{What test statistic can increase power in physics-motivated parameter regions?}
LRS is optimal for simple-vs-simple tests but is not guaranteed to be optimal for more general settings with high power against {\em all} alternatives. Hence, we propose an alternative test statistic (referred to as a {\em focused test statistic}; FTS) of the form 
\begin{equation}\label{eq:FTS}
    T_f(\D; \mu_0) = -2\log\left(\frac{p(\D; \mu_0)}{\int_{\Theta} p(\D; \mu) f(\mu) d\mu}\right),
\end{equation}
which allows the data analyst to focus the power of a test or confidence procedure to {\em specific} regions of the parameter space, as defined by the focus or weight function $f$. In Bayesian testing, this statistic is known as a Bayes factor. 
Whereas the idea of using the Bayes factor as a frequentist test statistic is not new \cite{good_bayesnon-bayes_1992,dalmasso_likelihood-free_2024,Fowlie03082023}, this work is the first to demonstrate that a test based on the Bayes factor has the potential to improve upon likelihood ratio tests in various particle physics settings. For simplicity, we will in our experiments use a Gaussian focus function $f(\mu)$ with mean $m$ and standard deviation $s$, truncated so that $f(\mu)=0$ for $\mu \notin \Theta$. The choice of a focus function may be determined through sensitivity studies on simulations, while the data is blinded. Alternatively, this choice may be guided by prior knowledge of $\mu_0$, for example, from an independent previous measurement or from theoretical considerations. The construction of an FTS confidence interval follows the procedure discussed above; see Figure~\ref{fig:test_statistics} (bottom). In complementary work, Ref~\cite{heinrich2022learning} discusses an improved treatment of nuisance parameters for non-asymptotic regimes.

\subsection{What are the statistical properties of our CIs?} \label{sec:statistical_properties}

Our proposed approach yields frequentist valid tests and confidence intervals, which have optimal power with respect to the focus function. More precisely:
\begin{itemize}
    \item {\bf Validity.} If the QR in Section~\ref{sec:quantile_regression} is perfectly estimated, then the confidence intervals $I(\mathcal{D})$ for $\mu$ are exact for every unknown parameter $\mu \in \Theta$: 
    \begin{equation}\label{eq:validity}
     \mathbb{P}_{\D|\mu}(\mu \in  I(\mathcal{D}))  = 1 - \alpha, \ \ \forall \mu \in \Theta.
      \end{equation}
    That is, we neither under- nor over-cover any subregions of $\Theta$.\footnote{Exact coverage is defined to hold ``instance-wise'' for every $\mu \in \Theta$, but not conditional on the data $\mathcal{D}$ or on subspaces of the data space. A discussion on scenarios with an under-fluctuation of the background is included in Sec.~\ref{sec:Conclusion}.} See Theorem 1 in \cite{dalmasso_likelihood-free_2024} for details and proof.
   \item {\bf Optimal power and shortest length.} Among all valid intervals, the FTS interval is the interval that minimizes the expected length $\mathbb{E} \left[|A(\D)|\right]$, where $|A(\D)|$ is the length of an interval $A$. The expectation or weighted average is computed with respect to the marginal distribution $p_X(x) = \int p(x;\mu) f(\mu) d\mu)$ induced by the focus (weight) function $f(\mu)$. See Appendix~\ref{app:optimality} for details and proof. 
    \item {\bf Focal point.}  By construction, our $(1-\alpha)100\%$ confidence intervals are nested and monotonically decreasing for increasing values of the error level $\alpha \in [0,1]$. The family of nested sets define a frequentist confidence distribution \cite{xie_confidence_2013}, where the $0 \%$ confidence interval---also known as the deepest point~\cite{liu_nonparametric_2022} or focal point~\cite{cortinovis_bayes-assisted_2025}---can be seen as a point estimator of $\mu$. The 0\% confidence interval coincides with Bayesian shrinkage estimators for likelihoods in the natural exponential family \cite{cortinovis_bayes-assisted_2025}. Such shrinkage estimators can offer improved performance and stability over Maximum Likelihood Estimate (MLE), especially for small data sets and high-dimensional problems (i.e., when a large number of parameters must be inferred). The ``shrinkage'' acts as a regularization and prevents models from overfitting the noise in the data, which is a common problem for MLE in high dimensions~\cite{stein_inadmissibility_1956}. While this work focuses on confidence interval estimation, if a physicist is instead interested in point estimation of one or a few parameters, the MLE is recommended.
    
\end{itemize}

\subsection{Are biased tests less desirable than UMPU tests?}
\label{sec:bias}
Bias in confidence intervals has a specific technical meaning: a $1-\alpha$ confidence interval $I(\D)$ for $\mu$ is \textit{biased} if there exists some $\mu'\neq \mu$ such that
\begin{equation*}
    \mathbb{P}_{\D\vert\mu}(\mu'\in I(\D)) > \mathbb{P}_{\D\vert\mu}(\mu\in I(\D)).
\end{equation*}
(See \cite[Definition 9.3.7]{casella_statistical_2024}.) Importantly, the presence of bias in this sense does not imply that the associated confidence intervals under- or over-cover; in particular, FTS sets remain valid. Historically, much effort has been devoted to the construction of uniformly most powerful unbiased (UMPU) tests. However, as noted by \cite{lehmann_testing_2005,suissa_are_1984}, biased tests may achieve greater power than UMPU tests for all alternatives of practical interest, at the cost of reduced power only for uninteresting alternatives or for a small subset of the alternative parameter space. Indeed, FTS deliberately introduces a small amount of bias (in parameter regions where the focus function has small weight) to increase power (generally in regions where the focus function has large weight), while retaining nominal coverage for every $\mu \in \Theta$ (Equation~\ref{eq:validity}). See Figure~\ref{fig:neyman_construction} in Appendix~\ref{sec:neyman_and_confidence_belts} for a detailed comparison between (both valid) LRT and FTS in terms of power and bias. Section~\ref{sec:LZLimits} illustrates with an example for searches that even if the focus is placed away from the true value, both the upper limits and lower limits provided by FTS outperform LRT across the interesting region of the parameter space. The introduction of this type of technical bias does not hinder a fully frequentist interpretation of the results and should not be mistaken for a constraint term sometimes introduced into the likelihood in particle physics, which can affect the validity of confidence intervals.

Many tests in nonparametric statistics are likewise biased to improve overall performance \cite{demidenko_approximations_2020,gordon_paradoxical_2010,murakami_unbiasedness_2023}. Finally, note that there is a direct analogy between tests and biased versus unbiased estimators (such as lasso versus ordinary least squares) in  statistics \cite{wainwright_high-dimensional_2019,wasserman_all_2004}.
Particle physics has a history of using biased tests such as the widely used CLs method~\cite{Read:2000ru}. In contrast to CLs, FTS confidence intervals guarantee correct coverage throughout the parameter space and a more principled rationale for assigning power, further discussed in Section~\ref{sec:choosingFocus}.

\subsection{How should one choose the focus function?}
\label{sec:choosingFocus}

Many models in statistics and statistical learning incorporate physics-motivated inductive biases to modify the power of the test in certain regions. The CLs method~\cite{Read:2000ru} in high-energy physics reduces the power of the test when the data should have no sensitivity, and the Feldman-Cousins procedure can incorporate physics constraints such as boundaries (for instance two sided boundaries for an angle parameter~\cite{schafer2003using}). The FTS offers an explicit mechanism for determining how statistical power varies across the parameter space.

While FTS has Bayesian roots, the focus function is not used for Bayesian inference: rather it serves as a {\em weight function} that determines where to allocate power of the test in the construction of frequentist confidence sets. The choice of the focus can therefore be optimized on independent simulated samples alongside other analysis decisions such as event selection, choice of observables or the size of bins used to construct template histograms. For a Higgs measurement, a focus function that minimizes the median $2\sigma$ CI on simulated samples (generated at the SM hypothesis) may be most appropriate. For searches, the community considers $3\sigma$ to be a benchmark for finding evidence of new physics and one may set that as a target for which to optimize the focus. Alternatively, one may optimize the focus near the default hypothesis that predicts no new physics, in order to quickly rule out alternate hypotheses. Physical constraints, such as two-sided boundaries for angle parameters, can straightforwardly be implemented using an appropriate focus function (e.g. a von Mises circular normal distribution), as can information from previous experiments. However, if data from previous experiments or auxiliary measurements are used to determine the focus function, that data should not be re-used in the final analysis  (a practice popularly referred to as ``double-dipping''~\cite{Kriegeskorte2009-uw}).

\section{Experiments and Results}
\label{sec:ExperimentsResults}

We highlight two applications -- a \textit{measurement} of the coupling of Higgs bosons to tau leptons and a \textit{search} for the existence of weakly interacting massive particles (WIMPs). The likelihoods are computed from bins of a histogram according to the counting experiment convention as described in Appendix~\ref{sec:experimental_setup}. We compare LRS confidence intervals with FTS where we use truncated Gaussian focus functions of various widths (both narrow and wide choices), to demonstrate the robustness of our qualitative conclusions to these choices as well as offering a rationale for making a specific choice in each study. To compute critical values, we adopt the procedure described in Section~\ref{sec:construction}.

\begin{figure}
    \centering
    \includegraphics{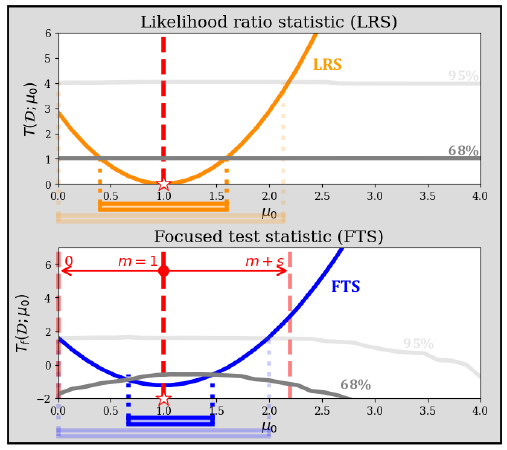}
    \caption{\textbf{Construction of confidence intervals.} (\textit{Top}) To construct a $(1-\alpha)100\%$ confidence interval by inverting the LRS statistic (yellow; Eq.~\ref{eq:LRS}), we estimate the critical values $C_{\mu_0}$ (grey) via quantile regression and retain those $\mu_0$ values for which the statistic falls below the critical value. Shown are the critical values for $68\%$ (dark grey) and $95\%$ (light grey) confidence levels, with corresponding intervals displayed below the figure (dark and light yellow, respectively). (\textit{Bottom}) The FTS statistic (blue; Eq.~\ref{eq:FTS}) for a focus function centered at $m=1$ and with width $s=1.2$ (dashed red) is compared against the critical values. Intervals at $68\%$ and $95\%$ confidence level are shown (dark and light blue, respectively). This figure corresponds to the Higgs measurement example using the Higgs mass observable, with the test statistics are evaluated on an Asimov data set~\cite{Cowan:2010js}}.
    \label{fig:test_statistics}
\end{figure}

\subsection{Measurement of a Higgs coupling via its signal strength}\label{sec:higgs} 

The goal of this case study is to measure the coupling of the Higgs boson to tau leptons~\cite{Aad:2015vsa}. This unknown quantity has a one-to-one correspondence with the signal strength in the Standard Model (SM) of particle physics and can therefore be posed as a signal strength measurement. The default hypothesis for the true signal strength $\mu^*=1$ corresponds to the SM prediction. For this study, we let $\Theta=[0, 10]$ to avoid right endpoint truncation bias in interval estimation, and we set the center of the focus function at $m=1$, the SM prediction. We define the wide focus width to be $s=2.4$, corresponding to the $2\sigma$ expected sensitivity of a prior measurement~\cite{ATLAS:2012gfw}; a narrow focus width of $s=1.2$ is also considered.

\subsubsection{Dataset} 

The dataset was simulated by the ATLAS experiment for the 2014 Higgs boson machine learning challenge (HiggsML)~\cite{HiggsMLOpenData,pmlr-v42-cowa14}. It is comprised of a mixture of events obtained via several Higgs decay and background processes. For further details on the challenge, refer to~\cite{pmlr-v42-cowa14}.

In order to show the robustness of the analysis to our experimental design choices, we examine two natural choices of observables to build template histograms for the counting experiment setup. The first observable is the reconstructed mass of the Higgs boson candidate as estimated via phase space integration; the second is the invariant mass of the hadronic tau and final-state lepton (``visible mass'' for short). 
Both observables are known to discriminate between signal and background, with the reconstructed mass expected to provide stronger constraints.

\subsubsection{Results}
When measuring the Higgs coupling with the reconstructed Higgs mass, FTS with either a wide ($s=2.4$) or narrow ($s=1.2$) focus gives 68\% confidence intervals\footnote{The 68\% and 95\% confidence intervals discussed throughout this Article correspond to $1 \sigma$ (68.27\%) and $2 \sigma$ (95.45\%) intervals, which is the convention in high-energy physics.} that have shorter median length than LRS. Figure~\ref{fig:median_lengths}(a) shows the median interval lengths as they vary with the true value of $\mu^*$ while the focus function remains unchanged, with center fixed at $m=1$. The improvement is robust to a misspecification of the focus function (meaning that the center of the focus function $m$ differs from the true value of the parameter $\mu^*$)
and the confidence intervals remain valid by construction. When $\mu^*=1$, the FTS intervals for a wide focus are about 13\% shorter than those obtained via LRS; see Table~\ref{tab:CI_summary}. With a narrow focus function, they are 21\% shorter than LRS intervals. As we find a marked improvement in median length from FTS-wide over LRS for a wider range of true signal strengths $\mu^*$ as compared to FTS-narrow, the wide focus function setting of Figure~\ref{fig:median_lengths}~(a, \textit{left}) is our preferred choice for this experiment.

Using the visible mass observable to constrain the parameter of interest, FTS shows a decrease in median length by 13\% with wide focus and 22\% with narrow focus. Table~\ref{tab:CI_summary} shows that the {\em absolute} reduction in median interval length of FTS over LRS is roughly the same at the 68\% and 95\% levels. However, since 95\% intervals are longer than 68\% intervals, this also implies that the proportional gain is smaller at 95\% than at 68\% (about a 6\% reduction for FTS-wide using the mass observable, 11\% for FTS-narrow; and a 6\% reduction for FTS-wide using visible mass, 10\% for FTS-narrow).

Estimation of critical values for both the LRS and FTS is accelerated by using quantile regression. With a low simulation budget of about $9{,}000$ pseudo-experiments (PEs), which for MC corresponds to 30 PEs per point times $300$ grid points, the mean-squared error (MSE) is improved by a factor of about 6 for LRS and 3 for FTS. See Figure~\ref{fig:critical_values} for an illustration of the relative speed-up.

\subsection{Search for WIMPs}\label{sec:lz}
Several experiments are dedicated to the search for dark matter candidates~\cite{Bozorgnia:2024pwk}. We design a data analysis scenario inspired by the search for WIMPs at the LZ experiment~\cite{LZ:2019sgr}. For our study, we construct a simplified dataset that mimics the signal and background densities, as well as the sample size, of the recent LZ dark matter search~\cite{LZ:2024zvo}. The default hypothesis is that these dark matter candidates do not exist, corresponding to $\mu^* = 0$; exclusion of this hypothesis would constitute a discovery. For this reason, our focus functions are centered at $m=0$. In this study, we let $\Theta=[0, 40]$ to avoid upper endpoint truncation bias in interval estimation; the focus function is thus truncated below at 0.

\subsubsection{Dataset}

We produce synthetic data mimicking the full dataset that passes selection cuts (as seen in~\cite[Fig.~3]{LZ:2024zvo}) relating densities of background and 40 GeV WIMP models (see Appendix~\ref{sec:LZ-inspired-dataset} for details). For consistency with the previous example, we use two-dimensional histograms for density estimation, although these can readily be replaced with an analytical model for the density, when they are available in dark matter experiments~\cite{Aalbers:2020iej}. The resulting 2D template histograms for the signal and background distributions are given in two observables, $S_1$ and $\log_{10}(S_2)$, representing the light from prompt vacuum ultraviolet
scintillation and delayed electroluminescence in the detector (see Fig.~\ref{fig:lz_histograms} in  Appendix~\ref{sec:LZ-inspired-dataset}). 
The simulated data are normalized such that when $\mu^*=1$, there are $1{,}200$ background events per $1$ signal event. The NEST simulator \cite{brodsky_nestcollaborationnest_2019} was used to generate the template histograms.
 
\subsubsection{Results: median CI lengths}

\begin{figure}[t!]
    \centering
    \resizebox{1.0\columnwidth}{!}{\includegraphics{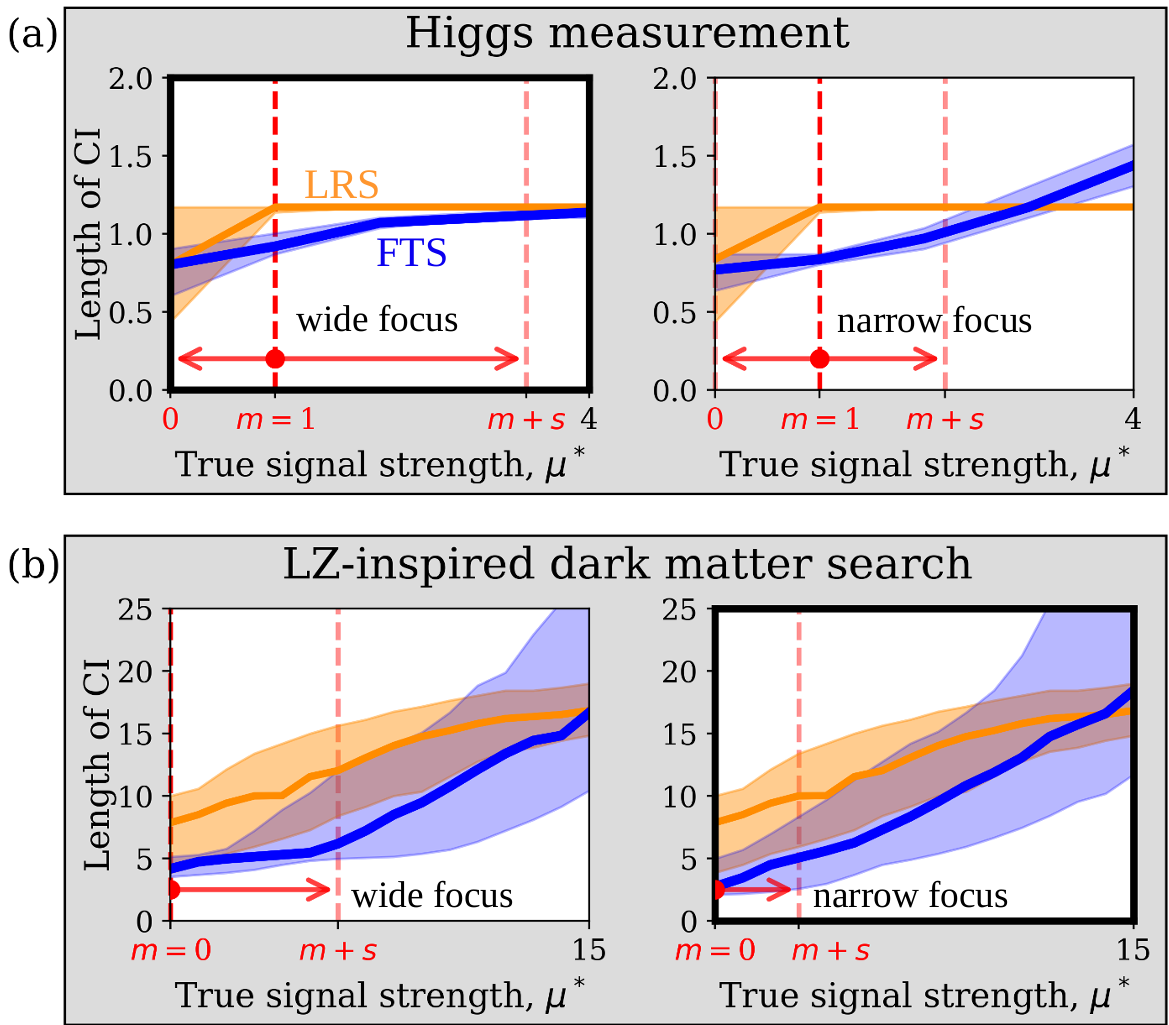}}
    \caption{\textbf{Lengths of confidence intervals.}  (a, \textit{left}) Results for the Higgs measurement with wide focus ($s=2.4$). The thick lines represent the median interval length with the uncertainty bands denoting 25\% and 75\% quantiles of the length distribution. FTS (blue) yields ~13\% shorter intervals than LRS (orange) near the focus center ($m = 1$).
    (a, \textit{right}) For a narrow focus ($s=1.2$), FTS intervals are about 25\% shorter. That advantage is maintained even at a modest distance from the center, e.g.~near $\mu^*=2$. (b, \textit{left-right}) For the {LZ-inspired search}, the intervals are shorter across the domain of our search. Near $\mu^*=0$, the FTS intervals  are about 35\% shorter with wide focus than LRS, and 22\% shorter with narrow focus. The two panels with black frames represent our focus choices in respective experiments.}
    \label{fig:median_lengths}
\end{figure}

\begin{figure}[t!]
    \centering
    \includegraphics{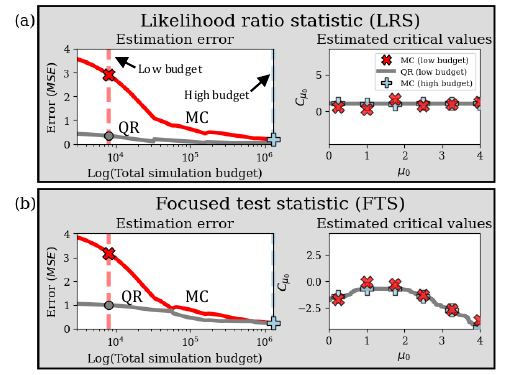} 
    \caption{\textbf{Critical values.} (a) We compare the mean-squared error (MSE) in the critical value estimates obtained via MC (red) and quantile regression (QR, grey) for LRS on a grid of $300$ evaluation points. For a low simulation budget of $9{,}000$ pseudo-experiments (PEs), MC estimates show high MSE whereas QR estimates (using the same number of PEs) are comparable to MC estimates with $1.35$ million PEs. The right column confirms that low-budget QR yields accurate estimates across the parameter space, unlike low-budget MC.
    (b) For FTS, QR is again more efficient. With $9{,}000$ PEs, QR matches the performance of MC with about $30{,}000$ PEs.}
    \label{fig:critical_values}
\end{figure}

In Figure~\ref{fig:median_lengths}b we show median confidence interval length in the LZ-inspired experiment as a function of the true signal strength. We find that FTS improves on LRS across the range of $0\leq \mu^*\leq 15$ with both wide ($s=6.0$) and narrow ($s=3.0$) focus functions. Table~\ref{tab:CI_summary} (bottom two rows)
shows that FTS with wide focus provides a median length $47\%$ shorter than LRS for a 68\% confidence level in the setting where $\mu^*=0$. With narrow focus, the FTS intervals are $65\%$ shorter. As a result, we would expect a true negative search to conclude sooner with our method relative to the LRS-based method, on average. Furthermore, FTS maintains an advantage over LRS when $\mu^*\neq 0$. When $\mu^*=1$, the median $68\%$ confidence interval lengths for FTS are still $45\%$ shorter than LRS intervals when using wide focus and $60\%$ with narrow focus. If WIMPs exist, our method would find evidence for them sooner by excluding the default hypothesis using less data. In the context, we recommend the narrow focus function since it demonstrates sensitivity to smaller signals. As in the Higgs case, the absolute gains in median interval length are similar for 68\% and 95\% confidence levels. For $\mu^*=0$, there is $11\%$ reduction for FTS-wide and for FTS-narrow; and for $\mu^*=1$, an $8\%$ reduction for FTS-wide and 10\% for FTS-narrow.

\begin{table}[h!]
    \caption{\textbf{Median confidence interval length (in units of $\mu$) at the 68\% (95\%) confidence level.}
    For the Higgs examples, the focus center is $m=1.0$; the focus widths are $s=2.4$ for FTS-wide, and $s=1.2$ for FTS-narrow. For LZ-inspired examples, the focus center is $m=0$; $s=6.0$ for FTS-wide, and $s=3.0$ for FTS-narrow. In each setting, the best result (the shortest median interval length) is bold-faced.}
    \label{tab:CI_summary}
    \resizebox{1.0\columnwidth}{!}{
    \begin{tabular}{c|c|c|c|c}
        \multicolumn{2}{c|}{Setting} & \multicolumn{3}{c}{Test statistic}\\\toprule
        Experiment & $\mu^*$ & LRS & FTS-wide & FTS-narrow \\\midrule
        Higgs (mass) & 1.0 & 1.08 (2.01) & 0.94 (1.89) & 0.85 (1.79) \\
        Higgs (vis.~mass) & 1.0 & 1.26 (2.31) & 1.10 (2.17) & 0.98 (2.08) \\
        \midrule
        LZ-inspired & 0.0 & 7.86 (12.10) & 4.17 (10.66) & 2.73 (10.74) \\
        LZ-inspired & 1.0 & 8.50 (13.47) & 4.73 (12.42) & 3.45 (12.10) \\
        \bottomrule
    \end{tabular}
    }
\end{table}

\subsubsection{Results: Limit setting}\label{sec:LZLimits}
\begin{figure}
    \resizebox{1.0\columnwidth}{!}{\includegraphics{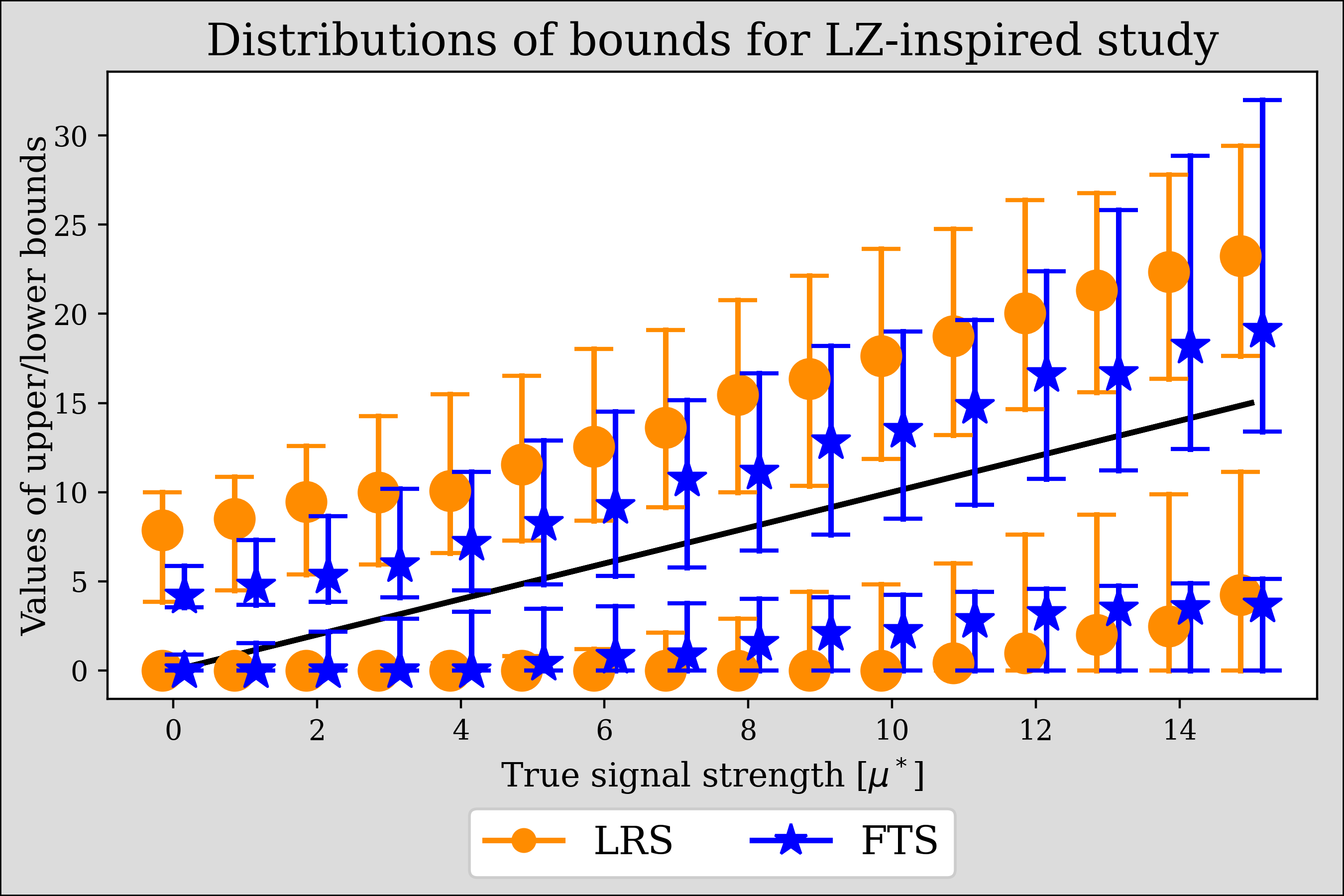}}
    \caption{Distributions of the upper and the lower bounds of 68\% confidence intervals for the LZ-inspired study, shown as box plots --- the median values are represented by circles (or stars) with the 25th and 75th percentiles of each distribution connected with a line. LRS (orange) is compared to FTS with wide focus (blue) over pseudo-experiments across different values of $\mu^*$ (x-axis). Both the median upper bound and the median lower bound are closer to the bisector for FTS than LRS, which is consistent with tighter parameter constraints. Note that a higher lower bound at small values of $\mu$ is especially desirable for searches, as it implies a higher chance of detecting weak signals.}
    \label{fig:lower_bounds}
\end{figure}

For the LZ-inspired search, it is interesting to study the upper limits when there is no signal (i.e. $\mu^*=0$) and lower limits when there is a signal (i.e. $\mu^*\neq 0$). Figure~\ref{fig:lower_bounds} compares the upper and lower bounds for these respective scenarios for LRS and FTS. These results for the no-signal scenario are summarized in Table~\ref{tab:LZUpperBounds} and results for a scenario where the signal does exist with $\mu^*=10$ is summarized in Table~\ref{tab:LZLowerBounds}. We find that when there is no signal, FTS is able to set tighter upper bounds, and when there is signal, FTS is able to set higher lower bounds compared to LRS. It is worth noting that focus is always placed at the default hypothesis of $\mu=0$ but FTS provides better lower limits when $\mu^*\neq0$. Applying a focus in a frequentist analysis therefore should not be confused with Bayesian inference.

\begin{table}
    \centering
    \begin{tabular}{c|c|c|c|c}
        $\mu^*$ & LRS & FTS, wide & FTS, narrow \\\hline
        0.0 & 7.86 (12.10) & 4.17 (10.66) & 2.73 (11.14) \\
    \end{tabular}
    \caption{Median upper bound for LZ-inspired experiment (in units of $\mu$) when there is no signal, $\mu^*=0$, at confidence level 68\% (95\%). The best result (the lowest median upper bound) is bold-faced.}
    \label{tab:LZUpperBounds}
\end{table}

\begin{table}
    \centering
    \begin{tabular}{c|c|c|c|c}
        $\mu^*$ & LRS & FTS, wide & FTS, narrow \\\hline
        10.0 & 0.0 (0.0) & 2.20 (0.0) & 1.28 (0.0) \\
    \end{tabular}
    \caption{Median lower bound for LZ-inspired experiment (in units of $\mu$) when the signal strength $\mu^*=10.0$, at confidence level 68\% (95\%). The best result (the highest median lower bound) is bold-faced.}
    \label{tab:LZLowerBounds}
\end{table}

\section{Discussion and Conclusion}
\label{sec:Conclusion}
In this Article, we propose the focus test statistic (FTS; Equation~\ref{eq:FTS}) as a drop-in replacement for the likelihood ratio statistic (LRS; Equation~\ref{eq:LRS}), which is widely used in high-energy physics for both parameter measurements and searches for new phenomena. These analyses rely on composite hypotheses, where the Neyman–Pearson lemma does not guarantee optimality of the LRS across the parameter space. In two proof-of-concept studies, a Higgs measurement at a collider experiment and a WIMP search in a dark matter experiment, we demonstrate that FTS can outperform LRS by focusing the power of the test in physics-motivated regions of the parameter space.
FTS with a narrow, well-specified focus function yields 22\% smaller 68\% confidence intervals than LRS for the Higgs measurement (when $\mu^*$ is at the SM point) and 65\% smaller intervals for the dark matter search (when there is no WIMP signal). For the dark matter search, we also find that FTS yields tighter upper bounds on the signal strength when the signal does not exist as well as higher lower bounds when it does, for a fixed amount of collected data. In our study, we take the variability between PEs into account and report the median interval length over PEs, together with the 25th and 75th percentiles of the length distribution.

While FTS has Bayesian roots, it is a fully frequentist method that provides valid confidence intervals. The FTS framework gives the scientist control over the regions of maximum sensitivity, whether chosen after optimization studies on independent simulations, informed by independent previous measurements (as in our Higgs example) or by theory, without affecting the validity of the confidence intervals. The focus function allows the analyst to leverage prior knowledge about reasonable parameter regions within a fully frequentist analysis. However, the data under analysis should not be used to determine the focus. For example, one can explore the impact of different focus functions on the sensitivity using simulated samples over a range of true values of the parameter (see Figure~\ref{fig:median_lengths}) to provide guidelines, if the simulations did not use the observed data. 
The general rule is that a narrow focus leads to larger gains in power around the center of mass of the focus function, but one then pays a price (as compared to a wider focus) in terms of decreased power {\em if} the focus center of mass differs from the true value of the parameter $\mu^*$, as illustrated by the top row of Figure~\ref{fig:median_lengths}. While FTS is biased in the technical sense, it is a deliberate choice to enhance the power of the test in physics-motivated regions without having to compromise on the coverage of the confidence intervals. The CLs method in particle physics also introduces a bias in order to adjust the power of the test in certain regions.

As an improvement to the core statistical methodology of the field, the FTS can enhance discovery potential and physics sensitivities across many experiments. Our case studies suggest that significant gains can be achieved when the signal-to-noise ratio is low and sample size is small, such as for dark matter searches, Higgs parameter measurements, and neutrino oscillation studies. Although not discussed in this Article, empty confidence sets are a known concern in HEP (e.g., under downward background fluctuations in counting experiments). For example, the $CL_s$ method~\cite{Read:2000ru} was developed to address this issue, and~\cite{Feldman:1997qc} provides another approach to avoiding empty confidence intervals. FTS could similarly be adapted to avoid empty confidence intervals, although such an adaptation may not preserve the optimality properties proven in Appendix~\ref{app:optimality}. Alternatively, one could report a singleton point estimate (e.g., the MLE, the focal point, or $\{0\}$) when the interval would otherwise be empty. We caution that analysts should not necessarily interpret or report an empty confidence interval as a meaningful exclusion of the signal model, especially when the experiment has little sensitivity to small signals. In such scenarios, we recommend the analyst report results together with the experiment's estimated sensitivity (see \cite[Section~VI]{Feldman:1997qc}).

We provide a convenient ML tool that builds confidence intervals by inverting hypothesis tests reliably and efficiently, with computational gains over traditional MC methods that are expected to scale with the dimensionality of the parameter space. The improved power of the FTS, enabled by the rapid building of the confidence intervals, can enhance the scientific discovery power of flagship measurements. Finally, our proposed method, which is here illustrated for histogram-based analyses, extends to unbinned, high-dimensional analyses with analytic expressions for the likelihood~\cite{Aalbers:2020iej} as well to as any other parametric~\cite{ATLAS:2018hxb} or nonparametric likelihood estimators~\cite{izbicki2014highdimensionaldensityratioestimation,Cranmer:2015bka,doi:10.1073/pnas.1912789117,ATLAS:2025clx,Ghosh:2025fma,dalmasso_likelihood-free_2024}.

\section*{Acknowledgments}

This project was initiated through discussions at the PHYSTAT workshops in Munich and London in 2024. The authors would like to thank the STAtistical Methods for the Physical Sciences (STAMPS) Research Center at Carnegie Mellon University for support. We are grateful to a reviewer for insightful comments, suggestions and discussions. We thank Lukas Heinrich, Mikael Kuusela, Ibles Olcina, David Rousseau, Larry Wasserman, and Callum Wilkinson for fruitful conversations. We thank Scott Kravitz for guidance on setting up the LZ-inspired dataset and pointing us to relevant resources. We also thank Scott Kravitz and Ben Nachman for feedback on the manuscript. AG and DW were supported by the DOE Office of Science. JC and ABL were supported in part by NSF DMS-2053804. RI is grateful for the financial support of FAPESP (grant 2023/07068-1) and CNPq (grants 305065/2023-8 and 403458/2025-0). This work used Bridges-2 at PSC through allocation MTH250002 from the Advanced Cyberinfrastructure Coordination Ecosystem: Services \& Support (ACCESS) program \cite{10.1145/3569951.3597559}, which is supported by U.S. National Science Foundation grants \#2138259, \#2138286, \#2138307, \#2137603, and \#2138296.

\section*{Data Availability Statement}

The code used to perform analysis and generate figures for this article can be found publicly available on GitHub at https://github.com/lee-group-cmu/focus-paper.

\clearpage

\appendix

\section{Experimental setup}
\label{sec:experimental_setup}
Over its run, an experiment measures events generated via signal and background physics processes. Large-scale simulations of these processes yield template histograms that define the expected counts of signal and background events as functions of relevant observables. We denote these histograms $\balpha(L)$ and $\bbeta(L)$ for the signal and background distributions, respectively, where $L$ is the luminosity (a measure of total count) of the experiment. 
We model the number of events that fall into each histogram bin as a Poisson distribution. That is, the probability of the number of events $\D=(N_{ 1}, N_{2}, \ldots, N_{d})$ in $d$ bins has the form
\begin{align}\label{eq:ATLAS-likelihood}
    p(\D ; \mu, \balpha, \bbeta) = \prod_{i=1}^d f(N_{ i}; \mu\alpha_i + \beta_i), 
\end{align}
where $\mu>0$ represents the signal strength, and $f(N; \lambda)$ denotes the probability mass function for a Poisson distribution with rate $\lambda$. We refer to each realization of $\D$ as a pseudo-experiment (PE). 

The above ``counting experiment'' model convention is widely used for making likelihood-free inference tractable in cases where rare or exotic events are mixed with a relatively high volume of background events~\cite{cowan1998statistical,casadei_reference_2012}. In this Letter, we follow the standard convention of histogram analyses for such counting experiments. 

\subsection{Implementation of test statistics}\label{sec:implementation}

We evaluate our test statistics on a grid of null parameters $\mu_0$ from $\Theta=[0, 10]$ in our main results for the Higgs experiment (Section~\ref{sec:higgs}), and from $[0, 40]$ for the LZ-inspired experiment (Section~\ref{sec:lz}). The grid is always of a resolution of at least 1 point per 0.04 units in $\mu$. In Figures~\ref{fig:test_statistics} and~\ref{fig:median_lengths}, we display results only on $[0, 4]$ for Higgs and $[0, 15]$ for LZ-inspired for the clear comparison of details.

For computing the LRS (Eq.~\ref{eq:LRS}), we replace $\Theta$ in the denominator with an extended parameter space, 
\begin{equation}\label{eq:extended_param_space}
    \widetilde \Theta =[ \min(\Theta)-1, \max(\Theta)+1].
\end{equation}
This adjustment mitigates the effect of the boundary on the distribution of the LRS itself, independent of the method used for critical value estimation. The boundary effect arises because Wilks' theorem, which guarantees that the critical values remain constant across the parameter space for LRS and large sample sizes (as in Figure~\ref{fig:test_statistics} top), only holds for {\em interior} points~\cite{Algeri:2019lah,Herold:2024enb}.
To ensure that the likelihood is well defined on $\mu<0$, we amend Eq.~\ref{eq:ATLAS-likelihood} to express 
\begin{align}\label{eq:ATLAS-likelihood-extended}
    p(\D ; \mu, \balpha, \bbeta) = \prod_{i=1}^d f(N_{ i}; \lvert\mu\alpha_i + \beta_i\rvert),
\end{align}
with rate parameter ensured to be positive. The supremum is computed using \textsc{scipy}'s implementation of the bounded Brent routine~\cite{2020SciPy-NMeth}.

\subsection{LZ-inspired dataset}
\label{sec:LZ-inspired-dataset}

\begin{figure}
    \resizebox{\columnwidth}{!}{
    \includegraphics{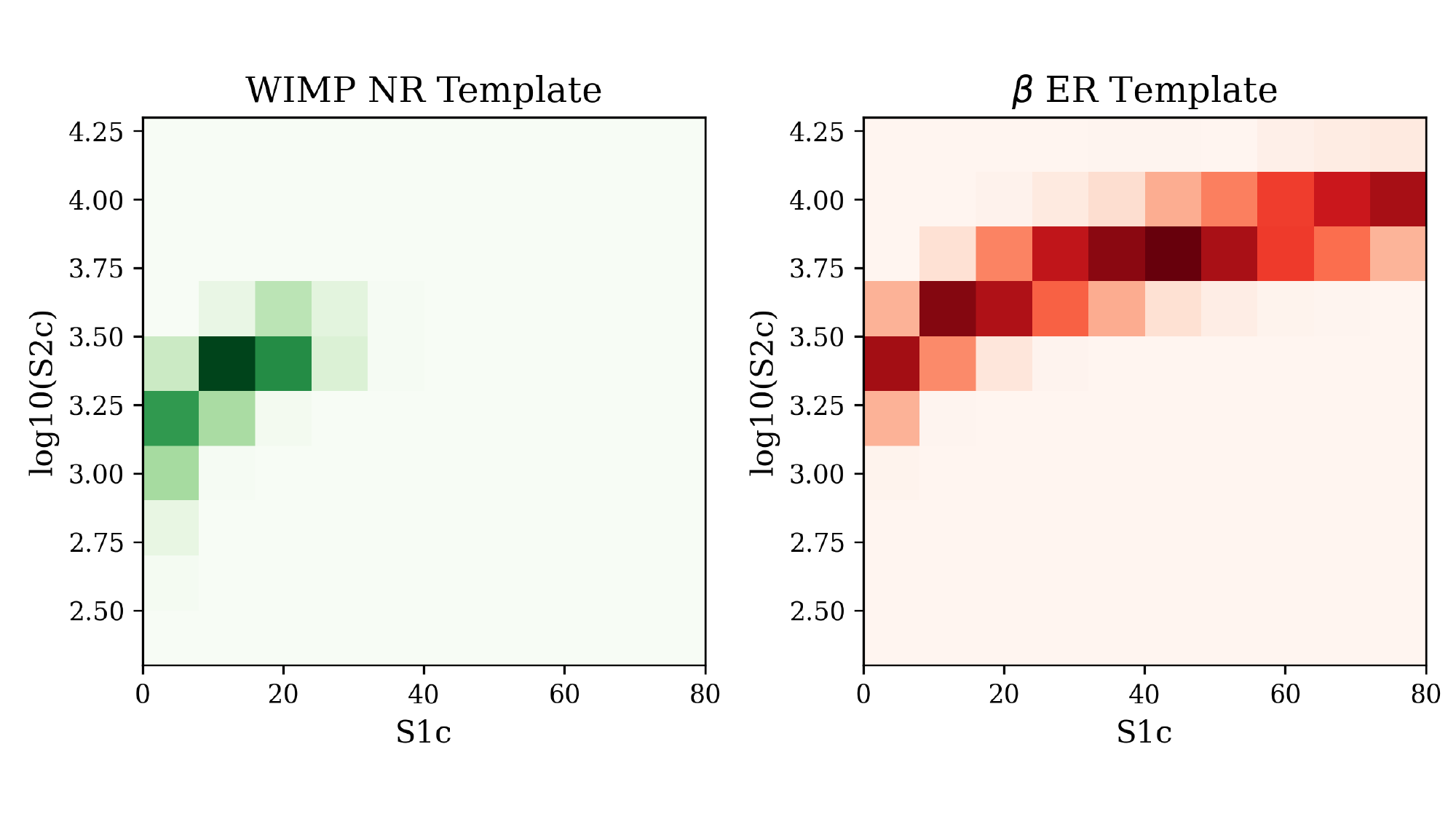}}
    \caption{Template 2D histograms for the LZ-inspired dataset, showing the signal (nuclear recoil, NR; \textit{left}) and background (electronic recoil, ER; \textit{right}). Each histogram has 10 bins along both axes.}
    \label{fig:lz_histograms}
\end{figure}

We construct a dataset with signal and background shapes visually similar to those in Fig.~3 of Ref.~\cite{LZ:2024zvo}. Template histograms are generated using the \textsc{nestpy} Python interface (\textsc{v2.0.4}) to the NEST simulator (\textsc{v2.4.0}). The \textsc{Lux\_Run03} detector binding is adapted for this study, with the default parameter values used to determine the background electronic recoil (ER) template histogram and with set values of $g_1=0.092$ and $g_1^{gas}=0.076$ for the signal nuclear recoil (NR) histogram. These values were chosen to visually match the 40 GeV WIMP signal distribution in Ref.~\cite{LZ:2024zvo}. The resulting histograms are shown in Fig.~\ref{fig:lz_histograms}.  More detailed reproduction of the LZ setup or the use of official LZ simulation samples is beyond the scope of this Article.

\section{Inverting hypothesis tests}
\label{sec:neyman_and_confidence_belts}

\subsection{Confidence belt}

A classical procedure for constructing a confidence set by inverting hypothesis tests, defined in Equation~\eqref{eq:CI}, is depicted in Ref~\cite{Feldman:1997qc} via a ``confidence belt'' plot. This plot graphically represents the procedure by displaying the acceptance regions for a discrete collection of tests concerning hypothesis $H_0:\mu=\mu_0$ versus $H_1:\mu\neq \mu_0$ for each $\mu_0$ in a grid.

\begin{figure*}
    \centering
    \resizebox{0.98\textwidth}{!}{\includegraphics{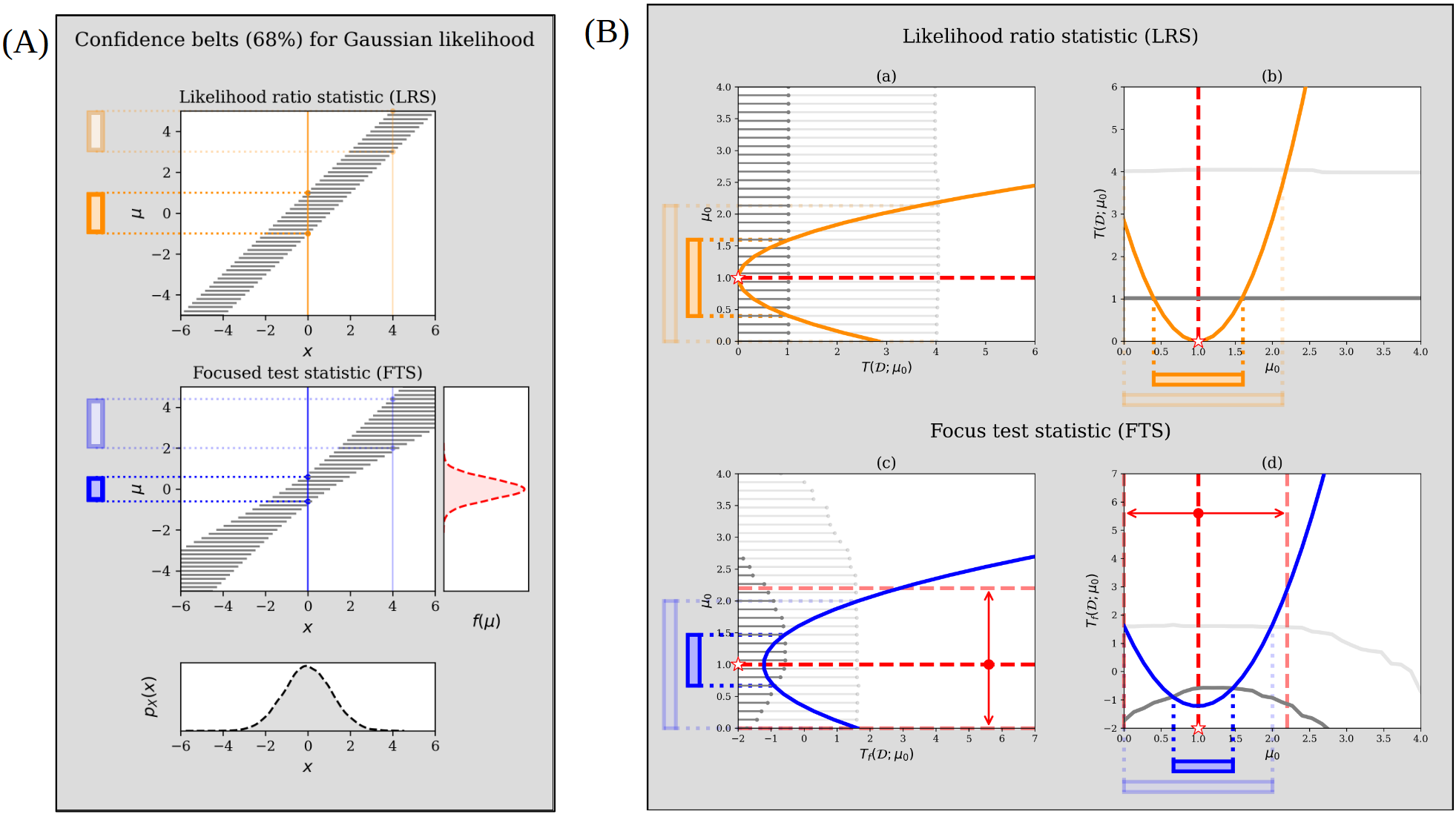}}
    \caption{\textbf{(A) Confidence belts for a Gaussian likelihood with unknown mean.} Two example confidence intervals are shown for the LRS (\textit{top}) and FTS (\textit{center}), using a Gaussian focus function (\textit{center-right}), for observations at $x=0$ and $x=4$. The interval obtained from FTS at $x=0$ (\textit{center}, solid blue) is shorter than that from LRS (\textit{top}, solid orange). At $x=4$, the FTS interval (\textit{center}, semi-transparent blue) is longer. The belts show that FTS intervals are shorter for values of $x$ where the marginal distribution $p_X(x) = \int p(x;\mu), f(d\mu)$ (\textit{bottom}), induced by the focus function, places most of its mass. This behavior is consistent with the theoretical result (Theorem~1 in Appendix~\ref{app:optimality}), which states that among all level-$\alpha$ confidence intervals $I(\mathcal{X})$, the FTS interval has the smallest expected length when the expectation is taken with respect to $p_X$. \textbf{(B) Visualization of confidence sets by inverting hypothesis tests.} The left column is adapted from \cite[Fig. 5(\textit{right})]{cranmer_practical_2015}. The right column is the same figure but rotated; in the main paper, we use the rotated visualization. See Figs.~\ref{fig:higgs_addl_examples} and~\ref{fig:lz_addl_examples} for more examples.}
    \label{fig:neyman_construction}
\end{figure*}

To illustrate, two examples of confidence belts are provided in Figure~\ref{fig:neyman_construction}(A) using the LRS (\textit{top}) and the FTS (\textit{center}). Both are for inference on the unknown mean $\mu$ of a Gaussian likelihood, $p(x;\mu)=\N(x; \mu, \sigma=1.0)$. The focus function used is $f(\mu)=\N(\mu; 0, \sigma=0.5)$. The figure shows that the width of the LRS confidence belt remains constant for this example, indicating constant expected interval size with respect to $x$, whereas the FTS confidence belt is narrower for $x$ values close to the marginal distribution of $X$, $p_X(x) = \int p(x;\mu) f(d\mu)$ (Figure~\ref{fig:neyman_construction}(A), \textit{bottom}). This result is as expected, because ``optimal power'' or ``minimum {\em average} size'' is defined with respect to the induced marginal distribution $p_X$; see Appendix~\ref{app:optimality} for theoretical proof.

\subsection{General visualization for inverting hypothesis tests}

\begin{figure*}
    \centering
    \resizebox{0.9\textwidth}{!}{\includegraphics{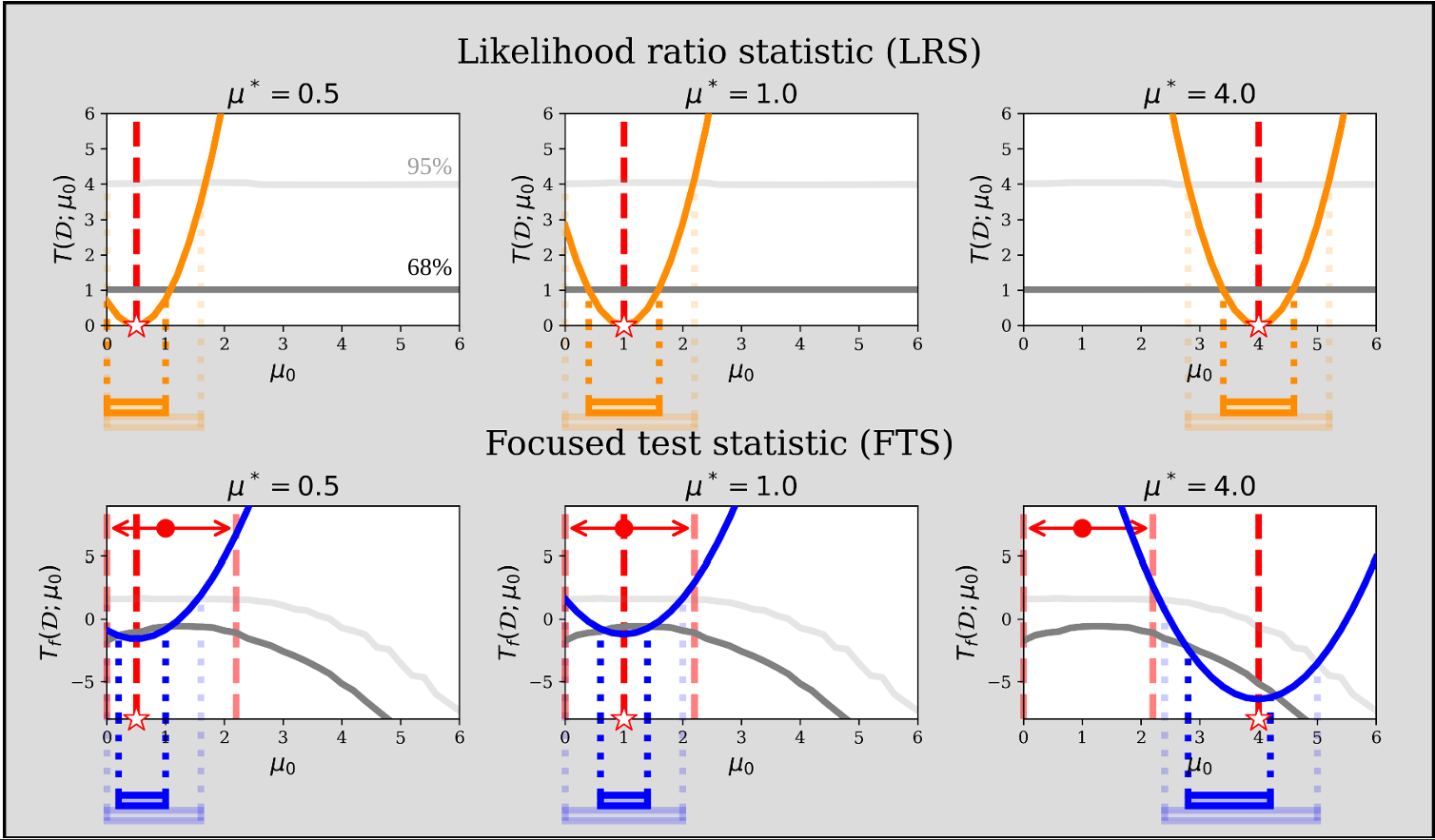}}
    \caption{\textbf{Confidence intervals for the Higgs example.} Each of LRS and FTS-wide are evaluated on the Asimov data set for true signal strengths $\mu^*=0.5, 1.0,$ and $4.0$. The length of the LRS intervals (\textit{top} row) does not visibly change as $\mu^*$ changes (e.g.~\textit{left} to \textit{right}) except due to truncation at the boundary, but the length of the FTS-wide intervals (\textit{bottom} row) does.}
    \label{fig:higgs_addl_examples}
\end{figure*}

\begin{figure*}
    \centering
    \resizebox{0.9\textwidth}{!}{\includegraphics{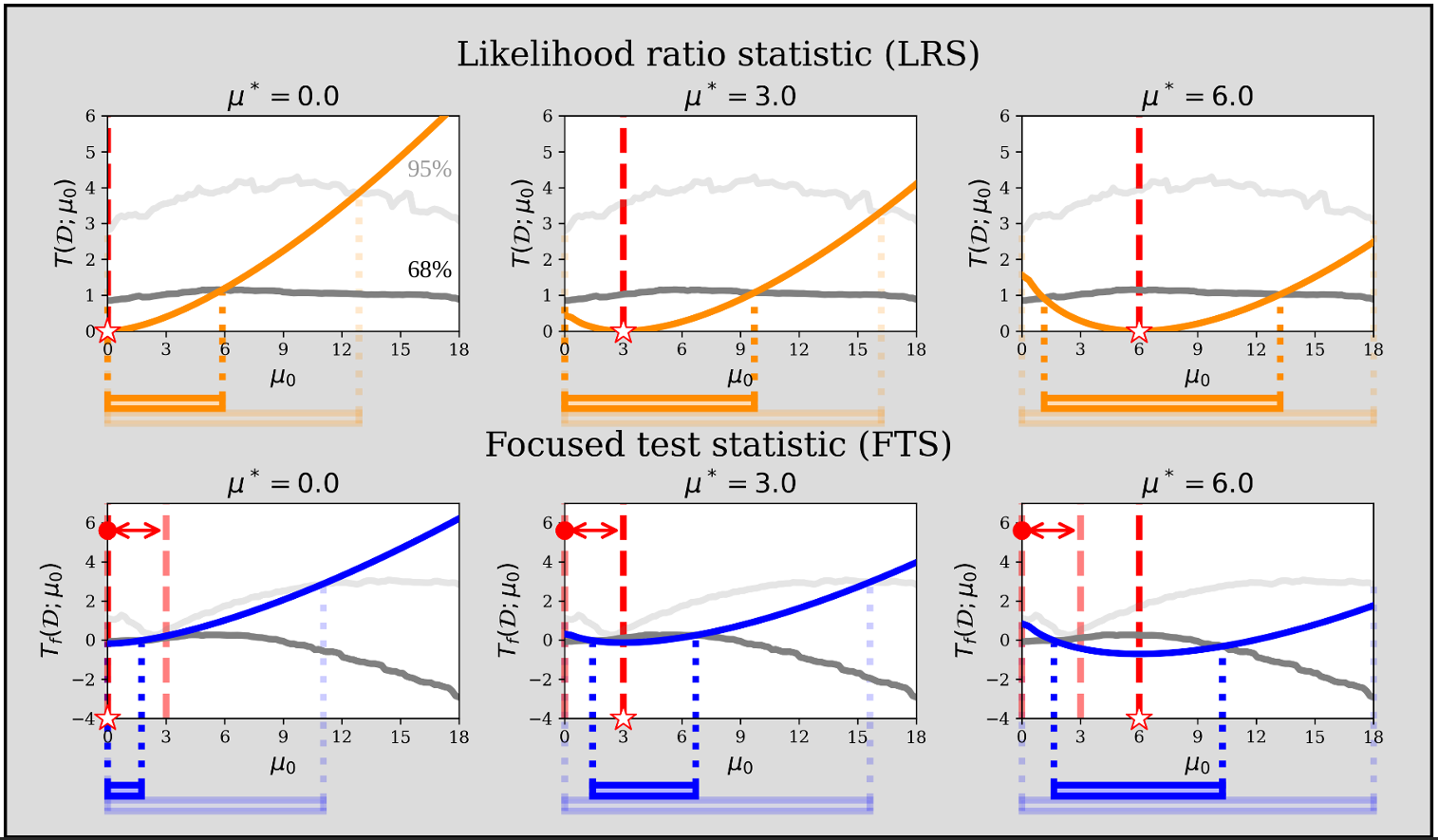}}
    \caption{\textbf{Confidence intervals for the LZ-inspired example.} Each of LRS and FTS-narrow are evaluated on the Asimov data set for true signal strengths $\mu^*=0.0, 3.0,$ and $6.0$. The critical value curves for the LRS for this data set (\textit{top} row; grey) are not flat, indicating that its sampling distribution is not well approximated by its asymptotic limit of $\chi^2_1$.}
    \label{fig:lz_addl_examples}
\end{figure*}

Ref.~\cite{cranmer_practical_2015} extends this visualization to more complex models with multi-channel data such as the binned likelihood in~\eqref{eq:ATLAS-likelihood} and we adopt this depiction in Figure~\ref{fig:neyman_construction}. The length of a confidence interval may depend on observed data. To see this explicitly for both of our experiments, we include a selection of examples of intervals for varied signal strengths comparing LRS with FTS with the focus choice highlighted in Figure~\ref{fig:median_lengths}. For the Higgs data set, we compare LRS with FTS-wide in Figure~\ref{fig:higgs_addl_examples}. For the LZ-inspired data set, we compare LRS with FTS-narrow in Figure~\ref{fig:lz_addl_examples}. Each example is evaluated on the Asimov data set\footnote{A specialized data set designed to represent the median expected sensitivity of an experiment \cite{Cowan:2010js}.} for the true signal strength $\mu^*$.

\section{Quantile regression}
\label{sec:critical_values}

To perform quantile regression, we generate pseudo-experiments to create a calibration dataset $\mathcal{T}=\{(\mu_1, \D_1),\ldots,(\mu_B, \D_B)\}$ according to Eq.~\ref{eq:ATLAS-likelihood} from $\mu_i\sim\text{Unif}(\widetilde{\Theta})$, where $\widetilde\Theta$ is given by Eq.~\ref{eq:extended_param_space}. For a code implementation of hypothesis test inversion using probabilistic regression along with examples for a range of test statistics, refer to \url{https://github.com/lee-group-cmu/lf2i}~\cite{dalmasso_likelihood-free_2024}.

\begin{algorithm}
\caption{Estimate critical values $C_{\mu_0}$ for a level-$\alpha$ test}
\label{alg:estimate_cutoffs} 
\KwData{test statistic $T(D;\mu)$; calibration data $\mathcal{T}=\{(\mu_1, D_1),\ldots,(\mu_B, D_B)\}$; quantile regression estimator; level $\alpha \in (0,1)$}
\KwResult{estimated critical values $\widehat{C}_{\mu_0}$ for all $\mu_0 \in \Theta$}

Set $\widetilde{\mathcal{T}}_{\rm cal} \gets \emptyset$\;

\For{$i \in \{1,\ldots, B\}$}{
    Compute test statistic $t_i \gets T(D_{i};\mu_i)$\;
    $\widetilde{\mathcal{T}}_{\rm cal} \gets \widetilde{\mathcal{T}}_{\rm cal} \cup  \{(\mu_i,t_{i})\}$\;
}

Use $\widetilde{\mathcal{T}}_{\rm cal}$ to learn the conditional quantile function $\widehat{C}_\mu = \widehat{F}_{T|\mu}^{-1}(\alpha | \mu)$ via quantile regression of $T$ on $\mu$\;

\Return{$\widehat{C}_{\mu_0}$}
\end{algorithm}

\section{Optimality of FTS confidence sets}
\label{app:optimality}

Let $\mathcal{D}\in\mathcal{X}$ denote the observed data and $\mu\in\Theta$ the parameter.
Fix a nonnegative focus (weight) function $f$ on $\Theta$ with $\int_\Theta f(\mu)\,d\mu=1$, and define the
\emph{focused (mixture) marginal}
\begin{equation}\label{eq:pf}
p_f(\mathcal{D}) \;:=\; \int_{\Theta} p(\mathcal{D};\mu)\, f(\mu)\, d\mu .
\end{equation}
Recall the focused test statistic (FTS) from Eq.~\eqref{eq:FTS}:
$$
T_f(\mathcal{D};\mu_0) \;=\; -2\log\!\left(\frac{p(\mathcal{D};\mu_0)}{p_f(\mathcal{D})}\right).
$$
For each $\mu_0$, let $C_{\mu_0}$ be a critical value satisfying
\begin{equation}\label{eq:crit}
\mathbb{P}_{\mathcal{D}\sim p(\cdot;\mu_0)}\!\left(T_f(\mathcal{D};\mu_0) < C_{\mu_0}\right)=1-\alpha .
\end{equation}
The Neyman confidence set associated to FTS is
\begin{equation}\label{eq:Balpha_focus}
B_\alpha(\mathcal{D})  
\;=\;\{\mu_0\in\Theta:\; T_f(\mathcal{D};\mu_0)<C_{\mu_0}\}.
\end{equation}

We will use the following version of Neyman-Pearson's lemma \cite{neyman_ix_1933}: 

\begin{lemma}[Neyman--Pearson]\label{lemma:NP_focus}
Let $\mu(z)$ and $\nu(z)$ be nonnegative integrable functions on a measurable space.
Fix $\alpha\in(0,1)$ and assume there exists $t$ such that
$A^*=\{z:\mu(z)/\nu(z)\ge t\}$ satisfies $\int_{A^*}\mu(z)\,dz=1-\alpha$.
Then $A^*$ solves
$$
\min_A \int_A \nu(z)\,dz \quad \text{subject to}\quad \int_A \mu(z)\,dz \ge 1-\alpha.
$$
\end{lemma}

The next theorem shows in which sense FTS is optimal. Different versions of this theorem have appeared in e.g. \cite{pratt_length_1961,yu_adaptive_2018,hoff_bayes-optimal_2023} for continuous $\Theta$, as well as \cite{sadinle_least_2019} when $\Theta$ is finite.

\begin{theorem}[Focused sets minimize expected length]\label{thm:best_region_focus}
Let $\mathcal{A}$ be the collection of measurable sets $A\subseteq \Theta\times\mathcal{X}$.
For $A\in\mathcal{A}$, write the $\mu$-section $A(\mathcal{D})=\{\mu:(\mu,\mathcal{D})\in A\}$ and its length
$|A(\mathcal{D})|=\int_{A(\mathcal{D})} d\mu$.
Consider the optimization problem
\begin{align*}
&\min_{A\in\mathcal{A}} \; \mathbb{E}_{\mathcal{D}\sim p_f}\!\left[|A(\mathcal{D})|\right]
\quad \text{subject to}\\ 
&\mathbb{P}_{\mathcal{D}\sim p(\cdot;\mu)}\big(\mu\in A(\mathcal{D})\big)\ge 1-\alpha,\ \ \forall \mu\in\Theta.    
\end{align*}
An optimizer $A^*$ is given by the FTS confidence set $B_\alpha(\mathcal{D})$ in Eq.~\eqref{eq:Balpha_focus}, that is, 
\begin{align*}
A^*(\mathcal{D})&=\Big\{\mu\in\Theta:\ \frac{p(\mathcal{D};\mu)}{p_f(\mathcal{D})}\ge t_\mu\Big\}
\;\\
&=\;\{\mu\in\Theta:\ T_f(\mathcal{D};\mu)<C_\mu\},    
\end{align*}
for suitable thresholds $t_\mu$ (or critical values $C_\mu$) chosen to satisfy the coverage constraints.
\end{theorem}

\begin{proof}
For $A\in\mathcal{A}$, let $A_\mu=\{\mathcal{D}:(\mu,\mathcal{D})\in A\}$ be the $\mathcal{D}$-section.
By Fubini,
\begin{align*}
\mathbb{E}_{\mathcal{D}\sim p_f}\!\left[|A(\mathcal{D})|\right]
&=\int_{\mathcal{X}}\!\left(\int_{A(\mathcal{D})}1\,d\mu\right)p_f(\mathcal{D})\,d\mathcal{D}\\
&=\int_{\Theta}\!\left(\int_{A_\mu} p_f(\mathcal{D})\,d\mathcal{D}\right)d\mu.    
\end{align*}
The constraints are $\int_{A_\mu} p(\mathcal{D};\mu)\,d\mathcal{D}\ge 1-\alpha$ for all $\mu$.
Hence the optimization problem decouples pointwise in $\mu$ into
$$
\min_{A_\mu}\int_{A_\mu} p_f(\mathcal{D})\,d\mathcal{D}
\quad \text{subject to}\quad
\int_{A_\mu} p(\mathcal{D};\mu)\,d\mathcal{D}\ge 1-\alpha.
$$
Applying Lemma~\ref{lemma:NP_focus} with $\mu(z)=p(z;\mu)$ and $\nu(z)=p_f(z)$ yields the optimal acceptance region
$$
A_\mu^*=\Big\{\mathcal{D}:\ \frac{p(\mathcal{D};\mu)}{p_f(\mathcal{D})}\ge t_\mu\Big\},
$$
with $t_\mu$ chosen so that
$\mathbb{P}_{\mathcal{D}\sim p(\cdot;\mu)}(\mathcal{D}\in A_\mu^*)=1-\alpha$.
Inverting acceptance regions gives
$$
A^*(\mathcal{D})=\{\mu:\ \mathcal{D}\in A_\mu^*\}
=\Big\{\mu:\ \frac{p(\mathcal{D};\mu)}{p_f(\mathcal{D})}\ge t_\mu\Big\}.
$$
Since $T_f(\mathcal{D};\mu)=-2\log\!\big(p(\mathcal{D};\mu)/p_f(\mathcal{D})\big)$,
this is equivalent to $\{\mu: T_f(\mathcal{D};\mu)<C_\mu\}$ for $C_\mu=-2\log t_\mu$,
which is exactly the FTS confidence set $B_\alpha(\mathcal{D})$ in \eqref{eq:Balpha_focus}.
\qed 
\end{proof}

The theorem shows that among all confidence sets that achieve at least $1-\alpha$ coverage \emph{for every} $\mu\in\Theta$,
the FTS  construction is optimal for average constraining power \emph{under the focus function}:
it minimizes the expected interval length when data are averaged according to the focused marginal
$p_f(\mathcal{D})=\int p(\mathcal{D};\mu)f(\mu)\,d\mu$.
In other words, choosing $f$ specifies where we want the procedure to be tight on average, and the FTS set is
the best (shortest on average) valid procedure for that choice.

\section{Rejection probability as a function of $\mu_0$} \label{sec:power_curves}

Our confidence procedure consists of a family of hypothesis tests, and a typical means of evaluating hypothesis tests is in terms of their statistical power via ``rejection probability functions.'' The rejection probability function $\rho(\mu)$ for the test concerning hypotheses $H_0:\mu=\mu_0$ and $H_1:\mu\neq\mu_0$ with test statistic $T$ is a function of $\mu$ and gives the probability under a fixed true signal strength $\mu^*$ that $\mu$ is included in the test's rejection region. The function is defined $\rho(\mu):=\mathbb{P}_{\D\sim p(\cdot; \mu^*)}(T(\D; \mu) > C_{\mu})$. In principle, an ideal rejection probability function would provide that $\rho(\mu^*)$ is equal to 0 (i.e.~the test never rejects the true signal strength) and that $\rho(\mu)$ is equal to 1 otherwise. In practice, hypothesis tests are typically designed so that $\rho(\mu^*) \leq \alpha$ for some prespecified Type I error level $\alpha$, and the goal for a test is for $\rho(\mu)$ to be as close to 1 as possible elsewhere.

\begin{figure*}
    \centering
    \resizebox{0.9\textwidth}{!}{    \includegraphics{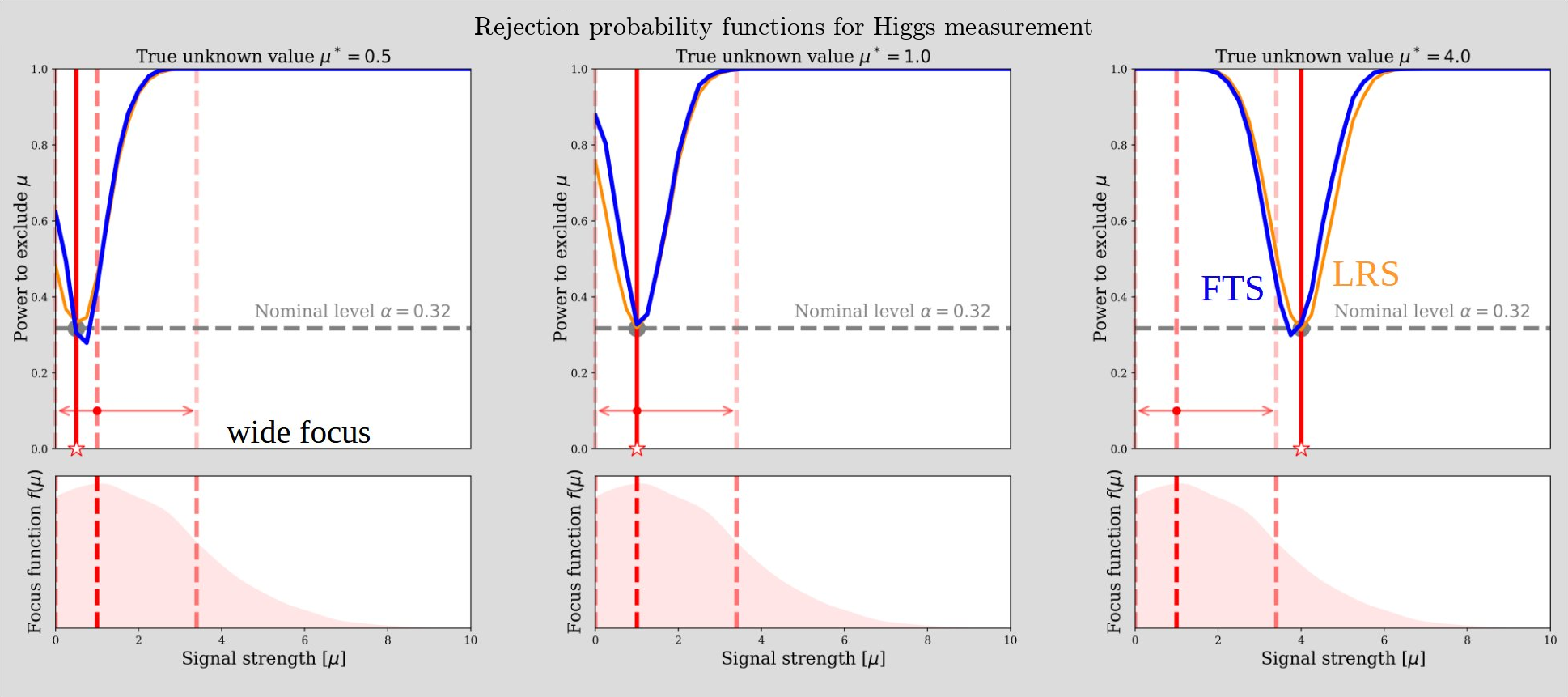}}
    \caption{\textbf{Rejection probability functions for the Higgs example.} LRS (blue) and FTS-wide (orange) are used to construct confidence intervals for $\D_1,\ldots,\D_{1,000}\sim p(\D;\mu^*)$ for $\mu^*=0.5, 1.0$, and $4.0$ (\textit{left}, \textit{center}, \textit{right}, respectively), and the resulting proportion of intervals retaining $\mu$ is plotted as a function of $\mu$.}
    \label{fig:higgs_power_curves}
\end{figure*}

In Figure~\ref{fig:higgs_power_curves}, we present rejection probability functions for the Higgs example of Section~\ref{sec:higgs} for both LRS and FTS-wide at the $1\sigma$ level. These curves correspond to true signal strengths $\mu^*=0.5, 1.0,$ and $4.0$. For example, Figure~\ref{fig:higgs_power_curves}(\textit{center}; $\mu^*=1.0$) gives a case where FTS' power curve is nowhere lower than LRS' and higher for some $\mu$. This is because $\mu^*$ is close to the center of the focus function. When the power curve for a test drops below the test level, then we say that that test is biased. However, this does not mean that the coverage of the confidence intervals is compromised. Figure~\ref{fig:higgs_power_curves}(\textit{right}; $\mu^*=4.0$) gives such an example. FTS' power curve falls below the nominal Type I error level here.

\begin{figure*}
    \centering
    \resizebox{0.9\textwidth}{!}{    \includegraphics{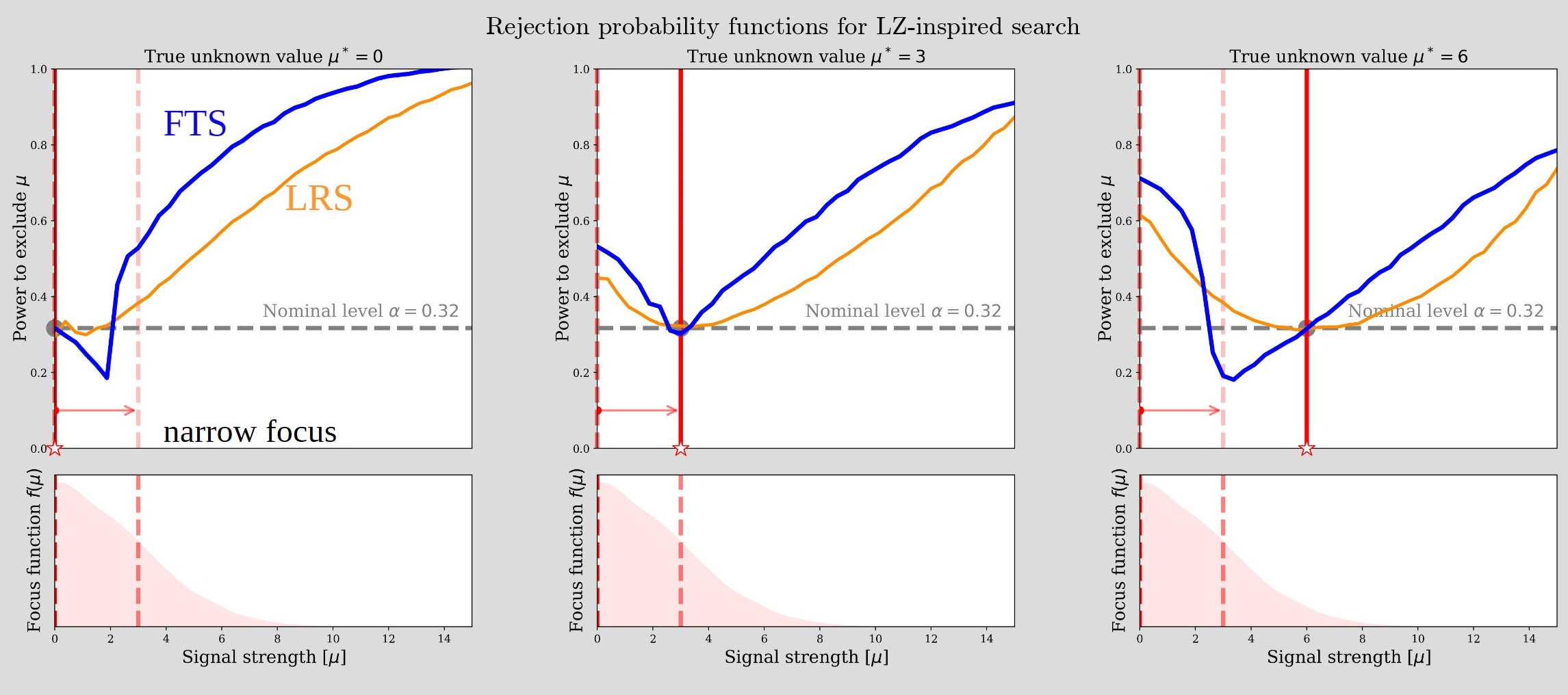}}
    \caption{\textbf{Rejection probability functions for the LZ-inspired example.} LRS (blue) and FTS-wide (orange) are used to construct confidence intervals for $\D_1,\ldots,\D_{1,000}\sim p(\D;\mu^*)$ for $\mu^*=0.0, 3.0$, and $6.0$ (\textit{left}, \textit{center}, \textit{right}, respectively), and the resulting proportion of intervals retaining $\mu$ is plotted as a function of $\mu$. In every case, FTS provides improved lower and upper bounds as seen in Figure~\ref{fig:lower_bounds}.}
    \label{fig:lz_power_curves}
\end{figure*}

Figure~\ref{fig:lz_power_curves} shows rejection probability functions for LRS and FTS-narrow for the LZ-inspired data set of Section~\ref{sec:lz}. A narrow focus results in a larger bias in FTS, as seen in Figure~\ref{fig:lz_power_curves}(\textit{left}; $\mu^*=0$), where $\mu=1.5$ is excluded from $I(\D)$ less than $20\%$, and (\textit{right}; $\mu^*=6$), where $\mu=2$ is also excluded from $I(\D)$ less than $20\%$ of the time. This low power to exclude some false signal strengths is incurred in exchange for high power to exclude others. Section~\ref{sec:LZLimits} shows that FTS intervals are more likely to exclude $\mu=0$ for smaller true signal strengths $\mu^*$ than LRS, and the upper limits from FTS are lower than those from LRS when $\mu^*=0$.

The end result of the chosen focus function is raising the lower bound for searches for small values of $\mu$. That is, despite some bias in the rejection probability function, we achieve all the desired physics outcomes for the problem at hand without in any way compromising the coverage of the confidence intervals.

\clearpage
%
\bibliographystyle{plain}
\bibliography{focus}
%

\end{document}